\documentclass{article}

\usepackage{amsmath,amsfonts,amssymb}
\usepackage{amsthm}
\usepackage{graphicx}
\usepackage{float}
\usepackage{caption}
\usepackage{subcaption}
\usepackage{geometry}
\usepackage{enumitem}
\usepackage{authblk}
\usepackage{microtype}
\usepackage{url}

\newtheorem{proposition}{Proposition}
\newtheorem{theorem}{Theorem}

\newtheorem{corollary}{Corollary}

\title{Absement: Quantitative Assessment of Metabolic Cost during Quasi-Isometric Muscle Loading}
\author{Serhii V. Marchenko\thanks{sergij.marchenko@onmedu.edu.ua}}
\affil{Department of Physiology, Pathophysiology, Medical Physics and Informatics, Odesa National Medical University, Odesa, Ukraine}
\date{}

\begin{document}

\maketitle

\begin{abstract}
Small deviations during nominally isometric loading can be separated into a sustained mean offset and fluctuations about that offset. We develop a local quasi-static model that connects these video-accessible kinematic quantities to metabolic energy. Muscle activation is eliminated through joint-moment equilibrium, and the resulting metabolic power is reduced to a smooth function of a locally invertible muscle-length coordinate. For the deviation \(x(t)=\ell(t)-\ell_0\), the reduced energy satisfies
\[
\mathcal{E}_{\mathrm{met}}[\ell]
=
P_0T
+
C_1\Delta\mathcal{A}_{\ell}
+
C_2\mathcal{M}_{2,\ell}
+
R_3,
\qquad
|R_3|
\leq
KT\|x\|_{L^\infty(0,T)}^3,
\]
where \(\Delta\mathcal{A}_{\ell}=\int_0^T x(t)\,dt\) is signed deviation absement and \(\mathcal{M}_{2,\ell}=\int_0^T x(t)^2\,dt\) is the second raw integral moment. Equivalently, if \(\mu_\ell\) and \(\sigma_\ell^2\) are the mean and variance of the observed length deviation, then \(\Delta\mathcal{A}_{\ell}=T\mu_\ell\) and \(\mathcal{M}_{2,\ell}=T(\mu_\ell^2+\sigma_\ell^2)\). A video-based protocol can therefore estimate the required predictors without differentiating the recorded trajectory. Within the autonomous quasi-static model, periodic variation does not require a separate cycle-specific predictor: its mean contributes through absement and its dispersion through the second moment. This moment-based protocol, rather than the Taylor expansion alone, provides the experimentally testable result: residual dependence on frequency after control for the first two moments would identify the limit of the quasi-static reduction.
\end{abstract}

\section{Introduction}

Maintaining an apparently static posture requires continuous muscular force production and postural correction. The resulting energetic expenditure depends not only on trial duration but also on the operating length of the active musculature, the required joint moment, the mean displacement from the target posture, and the magnitude of the accompanying fluctuations. Experimental studies of standing balance have demonstrated measurable energetic costs of stabilization and associations between metabolic expenditure and postural-control demands \cite{Houdijk2009,Houdijk2015,Monnard2017,MilesChan2017}. A duration-only descriptor therefore cannot distinguish trials that have the same duration but different mean muscle lengths or different distributions of postural deviation.

The metabolic cost of force generation is also not constant across mechanical conditions. Experimental and phenomenological models have related muscle energy expenditure to activation, force, contractile state, and muscle length \cite{Russ2002,Umberger2003,Bhargava2004}. More recent work indicates that rapid force production and time-varying isometric force can introduce additional energetic contributions \cite{VanDerZeeKuo2021,Muralidhar2025}, while fascicle operating length can affect the cost of cyclic force production \cite{Beck2022}. These findings motivate a formulation in which the local energetic response is derived from muscle equilibrium and metabolic-power laws instead of being assigned directly to a single kinematic index.

Within integral kinematics, \emph{absement} denotes the time integral of displacement \cite{Mann2006,MannJanzen2014}. For a selected muscle-length coordinate \(\ell(t)\) and reference value \(\ell_0\), the signed deviation absement is
\[
\Delta\mathcal{A}_{\ell}
=
\int_0^T\bigl(\ell(t)-\ell_0\bigr)\,dt
=
T\mu_\ell.
\]
It is therefore the duration-scaled first raw moment of the observed length deviation. Periodic or irregular corrections do not invalidate this quantity: a zero-mean component cancels from the first moment but remains present in the second and higher moments. This suggests a direct experimental strategy in which video-derived kinematics provide the empirical moments required by the energetic expansion.

The purpose of the present work is to derive this moment structure from a quasi-static muscle model and to identify the corresponding measurable predictors. We eliminate activation by the implicit function theorem, reduce instantaneous metabolic power to a smooth scalar function of an effective length coordinate, and obtain a controlled second-order energy expansion. The resulting model requires trial duration, mean length deviation, and deviation variance, rather than an explicit parameterization of every postural correction cycle.

The main contributions are as follows:
\begin{enumerate}[label=(\roman*)]
    \item a local reduction of quasi-static muscle equilibrium and metabolic power to a scalar energetic function of length;
    \item a rigorous second-order expansion in the first and second raw integral moments, with explicit coefficients and a uniform remainder estimate;
    \item an interpretation of periodic and irregular deviations through their empirical moment structure;
    \item a video-compatible identification protocol that reconstructs the relevant length coordinate before estimating its mean and variance;
    \item explicit conditions under which the reference posture is stationary or locally energy-minimizing.
\end{enumerate}

The result is deliberately local. It applies when the selected length coordinate is locally invertible with respect to joint angle, activation remains away from its bounds, quasi-static moment balance is appropriate, and metabolic power can be represented as a smooth instantaneous function of the reduced muscle state. Within this scope, absement is not introduced as an independent physiological law; it is the first raw integral moment generated by the Taylor expansion of the reduced energetic functional.

\section{Mathematical model and local reduction}\label{sec:model}

\subsection{Choice of coordinate and local geometry}

Consider a muscle group acting about one joint degree of freedom. Let
\(\theta(t)\) denote the joint angle, \(a(t)\in[0,1]\) the effective activation, and
\(\ell(t)\) a scalar length coordinate. The interpretation of \(\ell\) must be fixed before parameter estimation. It may denote a muscle--tendon path length inferred from joint kinematics and a musculoskeletal model, or an effective fascicle-length coordinate obtained from ultrasound. These quantities are not interchangeable.

For the present derivation, assume that the selected coordinate is locally represented by a \(C^3\) function
\[
\ell=\ell(\theta).
\]
Let \(\theta_0\) be the reference angle and define
\[
\ell_0:=\ell(\theta_0),
\qquad
s_0:=\ell_\theta(\theta_0),
\qquad
\kappa_0:=\ell_{\theta\theta}(\theta_0).
\]
The local kinematic expansion is
\begin{equation}\label{eq:kinematic-expansion}
\ell(\theta_0+y)
=
\ell_0+s_0y+\frac{1}{2}\kappa_0y^2+O(|y|^3).
\end{equation}
We assume
\begin{equation}\label{eq:s0-nonzero}
s_0\neq 0,
\end{equation}
so that \(\ell(\theta)\) has a local \(C^3\) inverse
\[
\theta=\Theta(\ell).
\]

The joint angle and the muscle moment-arm construction must be distinguished geometrically. Let \(\widehat{\mathbf e}_{\mathrm{h}}\) and \(\widehat{\mathbf e}_{\mathrm{f}}\) be unit vectors from the joint centre along the upper-arm and forearm axes, respectively. For the planar flexion shown in Figure~\ref{fig:model-schematic}, the unsigned joint angle is
\begin{equation}\label{eq:joint-angle-geometry}
\theta
=
\arccos\!\left(
\widehat{\mathbf e}_{\mathrm{h}}\mathbin{\cdot}
\widehat{\mathbf e}_{\mathrm{f}}
\right).
\end{equation}
By contrast, let the local line of action of the muscle force be
\[
\mathcal L_F(\theta)
=
\left\{
\mathbf r_F(\theta)+s\widehat{\mathbf t}_F(\theta)
:
s\in\mathbb R
\right\},
\]
where \(\widehat{\mathbf t}_F(\theta)\) is its unit tangent and \(\mathbf r_J\) is the joint-centre position. The orthogonal foot and moment arm are then
\begin{equation}\label{eq:moment-arm-projection}
\begin{aligned}
\mathbf r_\perp(\theta)
&=
\mathbf r_F(\theta)
+
\left[
\bigl(\mathbf r_J-\mathbf r_F(\theta)\bigr)
\mathbin{\cdot}
\widehat{\mathbf t}_F(\theta)
\right]
\widehat{\mathbf t}_F(\theta),\\
\rho(\theta)
&=
\left\|
\mathbf r_J-\mathbf r_\perp(\theta)
\right\|,
\qquad
\bigl(\mathbf r_J-\mathbf r_\perp(\theta)\bigr)
\mathbin{\cdot}
\widehat{\mathbf t}_F(\theta)
=0.
\end{aligned}
\end{equation}
Thus, the right angle in Figure~\ref{fig:model-schematic} does not define \(\theta\). It defines the perpendicular projection and therefore \(\rho(\theta)\) after the joint configuration has been set by \(\theta\). If \(\ell\) changes through the kinematic relation \(\ell=\ell(\theta)\), the line of action and its orthogonal foot are recomputed at the new configuration; the \(90^\circ\) condition is consequently preserved by definition, whereas neither the foot nor \(\rho\) is generally fixed.

The muscle moment arm is denoted separately by \(\rho(\theta)\). If \(\ell\) is the anatomical muscle--tendon path length, the usual geometric convention gives
\(\rho(\theta)=-\ell_\theta(\theta)\), up to the adopted sign convention. We do not impose this identity because \(\ell\) may instead be an effective or directly measured length coordinate. In particular, \(s_0\) and \(\rho(\theta_0)\) must not be represented by the same symbol unless their geometric relation has been explicitly imposed.

\subsection{Quasi-static equilibrium}

Let
\[
F_{\mathrm{tot}}(\ell,a)
\]
be the total muscle force entering the joint-moment balance. It may include active and passive components,
\[
F_{\mathrm{tot}}(\ell,a)
=
F_{\mathrm{act}}(\ell,a)+F_{\mathrm{pas}}(\ell).
\]
Let \(M_{\mathrm{ext}}(\theta)\) denote the external joint moment. Quasi-static equilibrium is expressed as
\begin{equation}\label{eq:equilibrium}
\rho(\theta)F_{\mathrm{tot}}\bigl(\ell(\theta),a\bigr)
=
M_{\mathrm{ext}}(\theta).
\end{equation}

In this scalar balance, \(\rho(\theta)\), \(F_{\mathrm{tot}}\), and \(M_{\mathrm{ext}}(\theta)\) denote nonnegative magnitudes, with \(M_{\mathrm{ext}}\) understood as the external moment opposing the muscle action. An equivalent signed formulation may be used, but it requires a signed moment arm and the same orientation convention for every moment throughout the derivation.

Define
\begin{equation}\label{eq:Q-def}
Q(\theta,a)
:=
\rho(\theta)F_{\mathrm{tot}}\bigl(\ell(\theta),a\bigr)
-
M_{\mathrm{ext}}(\theta).
\end{equation}

Let \((\theta_0,a_0)\), with \(a_0\in(0,1)\), satisfy
\[
Q(\theta_0,a_0)=0.
\]
Assume that \(Q\in C^3\) and
\begin{equation}\label{eq:Qa-nonzero}
Q_a(\theta_0,a_0)\neq 0.
\end{equation}
The implicit function theorem then provides a unique local \(C^3\) function
\[
a=a_\ast(\theta)
\]
such that
\[
a_\ast(\theta_0)=a_0,
\qquad
Q\bigl(\theta,a_\ast(\theta)\bigr)=0.
\]

Combining this function with the inverse kinematic map gives the equilibrium activation as a function of length:
\begin{equation}\label{eq:ahat}
\widehat a(\ell)
:=
a_\ast\bigl(\Theta(\ell)\bigr).
\end{equation}

\subsection{Metabolic-power functional}

Let
\[
\Pi(\ell,a)
\]
denote metabolic power. The analysis requires only that
\(\Pi\in C^3\) near \((\ell_0,a_0)\). Along a quasi-static equilibrium trajectory, define the reduced power
\begin{equation}\label{eq:reduced-power}
\varphi(\ell)
:=
\Pi\bigl(\ell,\widehat a(\ell)\bigr).
\end{equation}
The corresponding metabolic energy is
\begin{equation}\label{eq:energy-functional}
\mathcal{E}_{\mathrm{met}}[\ell]
=
\int_0^T\varphi\bigl(\ell(t)\bigr)\,dt.
\end{equation}

This formulation separates the force used in mechanical equilibrium from the metabolic-power law. For example, a possible effective specification is
\begin{equation}\label{eq:special-power-law}
\Pi(\ell,a)
=
\Pi_{\mathrm{b}}
+
\alpha a
+
\beta F_{\mathrm{act}}(\ell,a),
\qquad
\alpha,\beta>0,
\end{equation}
where \(\Pi_{\mathrm{b}}\) is a baseline power. The active force, rather than the total force including a passive component, appears in this particular metabolic approximation. Equation~\eqref{eq:special-power-law} is optional; the main result applies to the general function \(\Pi\).

For
\[
x(t):=\ell(t)-\ell_0,
\qquad
\varepsilon:=\|x\|_{L^\infty(0,T)},
\]
define the signed deviation absement and the second raw integral moment by
\begin{equation}\label{eq:absement}
\Delta\mathcal{A}_{\ell}
:=
\int_0^T x(t)\,dt
\end{equation}
and
\begin{equation}\label{eq:second-moment}
\mathcal{M}_{2,\ell}
:=
\int_0^T x(t)^2\,dt.
\end{equation}

\begin{figure}[H]
\centering
\includegraphics[width=\textwidth]{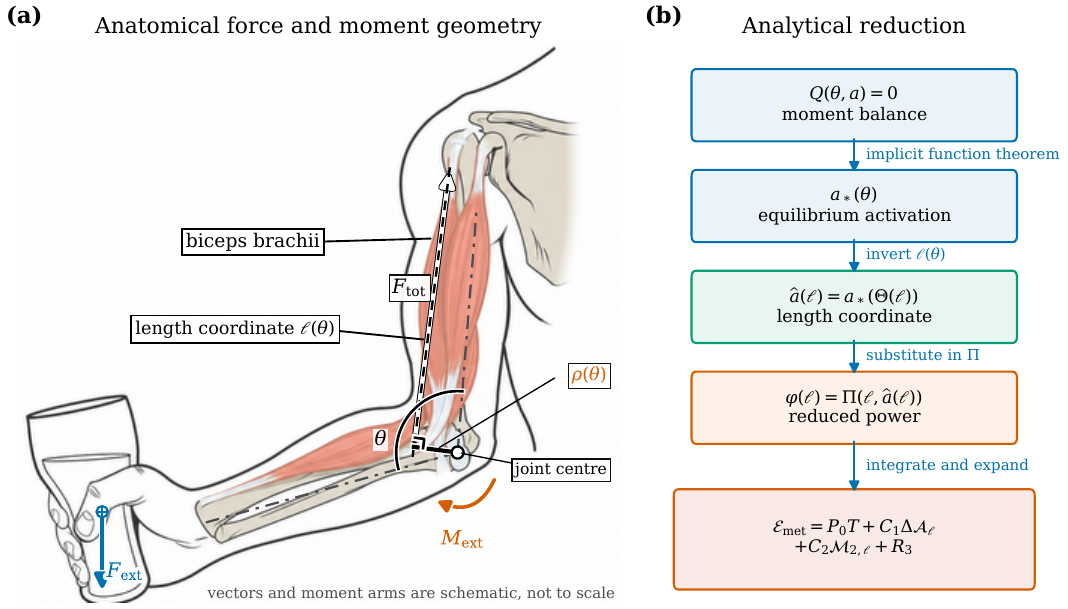}
\caption{Mechanical and analytical structure of the model. (a) Force and moment quantities are superimposed on an illustrative elbow-flexor system; all vectors and moment arms are schematic and not to scale. The two segment centrelines define the joint angle \(\theta\), which is the kinematic argument of both \(\ell(\theta)\) and \(\rho(\theta)\). The dashed line within the muscle-force vector represents the local line of action and the effective length direction. The perpendicular segment from the joint centre to this line defines \(\rho(\theta)\); its right-angle marker expresses Eq.~\eqref{eq:moment-arm-projection} and does not define \(\theta\). When the joint configuration changes, the line of action and its perpendicular foot are recomputed, so the projection remains orthogonal while \(\rho(\theta)\) may change. The marker on the glass identifies the assumed load centre of mass from which \(F_{\mathrm{ext}}\) acts. The anatomical base was adapted from OpenStax College \cite{OpenStaxBiceps2013} by cropping, removal of the original labels and arrows, conversion to grayscale, fading, and addition of the present annotations. (b) Analytical reduction from moment balance to the scalar energetic functional. The coordinate derivative \(s_0=\ell_\theta(\theta_0)\) is not identified with \(\rho(\theta_0)\) unless the anatomical path-length convention is imposed.}
\label{fig:model-schematic}
\end{figure}

Figure~\ref{fig:model-schematic} summarizes the sequence of reductions used below. The construction first eliminates activation through moment balance and only then expands the reduced power in the selected length coordinate. The biceps illustration supplies anatomical orientation only: no geometric quantity in the derivation is estimated from the drawing, and the same reduction applies to another muscle group whenever the stated one-coordinate assumptions hold.

\section{Local expansion of metabolic energy}

The preceding reduction leaves one experimentally relevant question: which features of an observed length trajectory can affect total metabolic energy near the reference posture? The answer does not require an independent energetic parameter for every postural correction. Because the reduced power is locally a scalar function of length, its first two derivatives select the first two raw integral moments of the recorded deviation.

The following theorem makes this statement precise. Its linear term measures the accumulated directional offset from the reference, whereas its quadratic term combines the squared mean offset with dispersion about that mean. The theorem therefore supplies the mathematical link between the reduced muscle model and the video-based predictors introduced in Section~\ref{sec:video-identification}.

\begin{theorem}[Local absement expansion]\label{thm:main}
Suppose that \(\ell(\theta)\), \(Q(\theta,a)\), and
\(\Pi(\ell,a)\) satisfy the regularity and non-degeneracy assumptions stated above. Then, for every admissible trajectory with sufficiently small
\(\varepsilon=\|\ell-\ell_0\|_{L^\infty(0,T)}\),
\begin{equation}\label{eq:main-expansion}
\mathcal{E}_{\mathrm{met}}[\ell]
=
P_0T
+
C_1\Delta\mathcal{A}_{\ell}
+
C_2\mathcal{M}_{2,\ell}
+
R_3[\ell],
\end{equation}
where
\begin{equation}\label{eq:main-coefficients}
P_0=\varphi(\ell_0),
\qquad
C_1=\varphi'(\ell_0),
\qquad
C_2=\frac{1}{2}\varphi''(\ell_0).
\end{equation}
Moreover,
\begin{equation}\label{eq:remainder-bound}
|R_3[\ell]|
\leq
\frac{T}{6}
\sup_{|z-\ell_0|\leq\varepsilon}
|\varphi'''(z)|
\,\varepsilon^3.
\end{equation}
\end{theorem}

\begin{proof}
Taylor's theorem applied pointwise to the reduced power gives
\[
\varphi(\ell_0+x)
=
\varphi(\ell_0)
+
\varphi'(\ell_0)x
+
\frac{1}{2}\varphi''(\ell_0)x^2
+
r_3(x),
\]
where
\[
|r_3(x)|
\leq
\frac{1}{6}
\sup_{|z-\ell_0|\leq\varepsilon}
|\varphi'''(z)|
\,|x|^3.
\]
Substitution of \(x=x(t)\) and integration over \([0,T]\) yield
\[
\begin{aligned}
\mathcal{E}_{\mathrm{met}}[\ell]
&=
\varphi(\ell_0)T
+
\varphi'(\ell_0)\int_0^T x(t)\,dt \\
&\quad
+
\frac{1}{2}\varphi''(\ell_0)
\int_0^T x(t)^2\,dt
+
\int_0^T r_3\bigl(x(t)\bigr)\,dt.
\end{aligned}
\]
Equations~\eqref{eq:absement}--\eqref{eq:main-coefficients} give
\eqref{eq:main-expansion}. Finally,
\[
\left|
\int_0^T r_3\bigl(x(t)\bigr)\,dt
\right|
\leq
\frac{T}{6}
\sup_{|z-\ell_0|\leq\varepsilon}
|\varphi'''(z)|
\,\varepsilon^3,
\]
which proves \eqref{eq:remainder-bound}.
\end{proof}

\begin{corollary}[First variation]\label{cor:first-variation}
For any bounded perturbation \(h\),
\begin{equation}\label{eq:first-variation}
D\mathcal{E}_{\mathrm{met}}(\ell_0)[h]
=
C_1\int_0^T h(t)\,dt.
\end{equation}
Consequently, if \(C_1\neq 0\), deviation absement is the only integral coordinate generated by the first-order term of the autonomous local functional~\eqref{eq:energy-functional}. If \(C_1=0\), the first variation vanishes for every perturbation and the leading nonconstant contribution is of second order.
\end{corollary}

\begin{proof}
For \(\ell_\eta(t)=\ell_0+\eta h(t)\),
\[
\begin{aligned}
D\mathcal{E}_{\mathrm{met}}(\ell_0)[h]
&=
\left.
\frac{d}{d\eta}
\right|_{\eta=0}
\int_0^T
\varphi\bigl(\ell_0+\eta h(t)\bigr)\,dt \\
&=
\varphi'(\ell_0)\int_0^T h(t)\,dt
=
C_1\int_0^T h(t)\,dt.
\end{aligned}
\]
\end{proof}

\begin{proposition}[Rearrangement invariance]\label{prop:rearrangement}
Let \(\ell\) be an admissible trajectory and let \(\pi:[0,T]\to[0,T]\) be a measurable measure-preserving transformation such that \(\ell\circ\pi\) is also admissible. Then
\[
\mathcal{E}_{\mathrm{met}}[\ell\circ\pi]
=
\mathcal{E}_{\mathrm{met}}[\ell],
\]
and the quantities
\(\Delta\mathcal{A}_{\ell}\) and \(\mathcal{M}_{2,\ell}\) are also unchanged by the transformation.
\end{proposition}

\begin{proof}
For every integrable scalar function \(g\), measure preservation gives
\[
\int_0^T g\bigl(\ell(\pi(t))\bigr)\,dt
=
\int_0^T g\bigl(\ell(t)\bigr)\,dt.
\]
The result follows by choosing
\(g=\varphi\), \(g(z)=z-\ell_0\), and
\(g(z)=(z-\ell_0)^2\).
\end{proof}

Proposition~\ref{prop:rearrangement} identifies a useful invariance of the quasi-static model. Profiles that spend the same fraction of the trial at the same length values have the same prediction, even if those values occur in a different order. Consequently, periodic or irregular corrections do not require separate cycle-specific coefficients when their empirical length distributions agree. The same invariance also marks the limit of the autonomous reduction: a systematic dependence on frequency after controlling for the first two moments would indicate the need for rate-dependent variables.

\subsection{Origin of the effective coefficients}

The coefficients \(C_1\) and \(C_2\) should not be interpreted as universal metabolic constants. Through the equilibrium activation \(a_\ast(\theta)\), they combine local muscle geometry, external moment, force--length properties, and derivatives of the metabolic-power law. For the interpretation and experimental regression developed below, only their aggregate values are required.

The complete implicit-differentiation and chain-rule expressions are derived in Appendix~\ref{app:explicit-coefficients}. That derivation also retains the quadratic correction generated by the nonlinear inverse relation between joint angle and muscle length.

\section{Interpretation and experimental identification}

\subsection{Mean displacement and variance}

A nominally static trial usually contains two geometrically distinct components. The participant may remain systematically displaced from the reference posture, and the observed trajectory may fluctuate around that displaced position. The first component changes the signed absement; the second is invisible to signed absement when its mean is zero but remains present in the quadratic contribution.

These components are summarized by
\begin{align}
\mu_\ell
&:=
\frac{1}{T}
\int_0^T x(t)\,dt
=
\frac{\Delta\mathcal{A}_{\ell}}{T},
\label{eq:mean-deviation}\\
\sigma_\ell^2
&:=
\frac{1}{T}
\int_0^T
\bigl(x(t)-\mu_\ell\bigr)^2\,dt,
\label{eq:variance}\\
\mathcal{M}_{2,\ell}
&=
T\bigl(\mu_\ell^2+\sigma_\ell^2\bigr).
\label{eq:raw-moment-decomposition}
\end{align}
Thus, the first raw moment records direction and mean offset, whereas the second raw moment combines that offset with dispersion.

\begin{figure}[H]
\centering
\includegraphics[width=\textwidth]{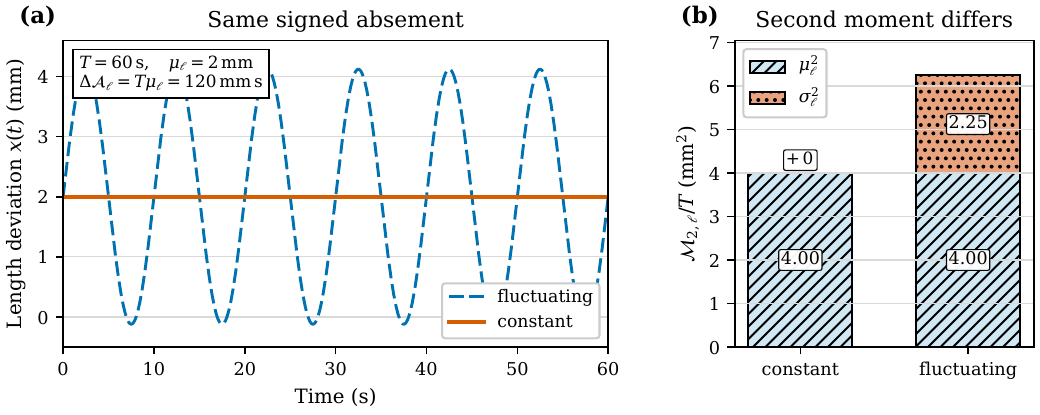}
\caption{How periodic variation enters the moment expansion. (a) Both constructed trajectories have \(T=60\,\mathrm{s}\), mean deviation \(\mu_\ell=2\,\mathrm{mm}\), and signed absement \(\Delta\mathcal{A}_\ell=120\,\mathrm{mm\,s}\). The fluctuating trajectory contains six complete sinusoidal cycles with \(\sigma_\ell=1.5\,\mathrm{mm}\). Its periodic component has zero mean and therefore does not change the signed absement. (b) The same component increases \(\mathcal{M}_{2,\ell}/T\) from \(4.00\) to \(6.25\,\mathrm{mm^2}\). Within the quasi-static model, the mean offset therefore supplies the first-order predictor and the dispersion supplies the additional second-order predictor. Fill patterns and line styles preserve the comparison in grayscale.}
\label{fig:controlled-trajectories}
\end{figure}

Figure~\ref{fig:controlled-trajectories} illustrates why periodicity is not, by itself, an obstacle to the proposed description. Over complete cycles, the positive and negative parts of the sinusoidal correction cancel in the first moment. They do not disappear physically or mathematically; their squared amplitudes survive in the second moment. For
\[
x_{\mathrm{per}}(t)
=
\mu_\ell+\sqrt{2}\sigma_\ell\sin(\omega t),
\qquad
\omega T=2\pi N,
\]
the two relevant integrals reduce to
\begin{equation}\label{eq:periodic-moments}
\Delta\mathcal{A}_{\ell}
=
T\mu_\ell,
\qquad
\mathcal{M}_{2,\ell}
=
T\bigl(\mu_\ell^2+\sigma_\ell^2\bigr).
\end{equation}
Boundary corrections for incomplete cycles are derived in Appendix~\ref{app:video-moments}.

The energetic prediction can consequently be written as
\begin{equation}\label{eq:mean-variance-energy}
\frac{\mathcal{E}_{\mathrm{met}}}{T}
=
P_0
+
C_1\mu_\ell
+
C_2
\bigl(
\mu_\ell^2+\sigma_\ell^2
\bigr)
+
\frac{R_3}{T}.
\end{equation}
This is the central form for experimental interpretation. The contribution \(C_1\mu_\ell\) is the power-normalized absement correction. The term \(C_2\mu_\ell^2\) accounts for the quadratic effect of a sustained offset, and \(C_2\sigma_\ell^2\) accounts for dispersion about that offset.

More generally, the autonomous model depends on how much of the trial is spent at each length value, not on the order in which those values occur. This statement is formalized through the empirical occupation measure in Appendix~\ref{app:video-moments}. It explains why periodic, phase-shifted, or irregular profiles with the same distribution of length values have the same quasi-static prediction.

The signed nature of absement must nevertheless be retained. Replacing it by an absolute-value integral would suppress directional cancellation and define a different, non-differentiable model. Likewise, frequency independence should not be assumed outside the quasi-static range: a residual frequency effect after controlling for \(\mu_\ell\) and \(\sigma_\ell^2\) would indicate the need for rate-dependent terms.

\subsection{Stationarity and local energetic optimality}

The reference posture is stationary with respect to arbitrary small length perturbations if and only if
\begin{equation}\label{eq:stationarity}
C_1=\varphi'(\ell_0)=0.
\end{equation}
If, in addition,
\begin{equation}\label{eq:positive-curvature}
C_2=\frac{1}{2}\varphi''(\ell_0)>0,
\end{equation}
then \(\ell_0\) is a strict local minimum of the reduced instantaneous power and the leading energetic penalty is quadratic:
\[
\mathcal{E}_{\mathrm{met}}[\ell]-P_0T
=
C_2\mathcal{M}_{2,\ell}
+
O(T\varepsilon^3).
\]
If \(C_1\neq0\), the selected reference posture is not an unconstrained energetic optimum in the chosen coordinate. It may nevertheless be imposed by task geometry, stability, comfort, or an external performance constraint.

\begin{figure}[H]
    \centering
    \includegraphics[width=\textwidth]{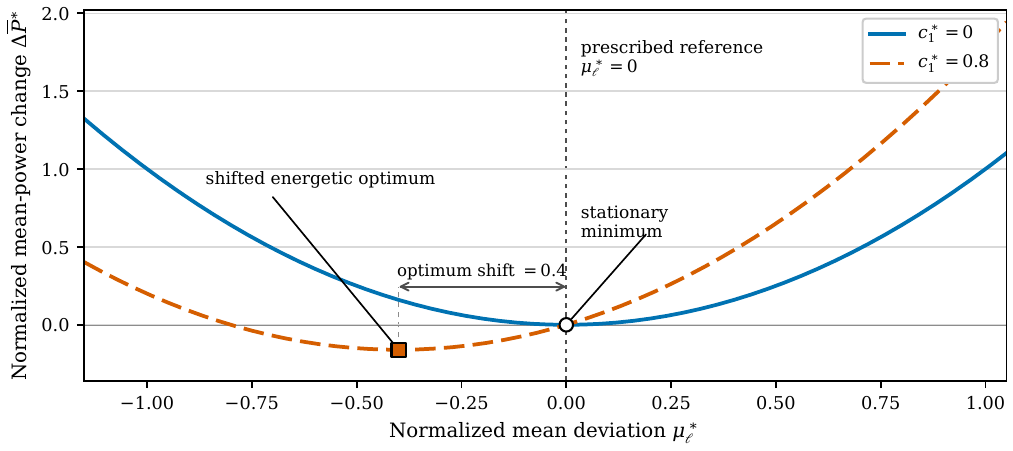}
    \caption{Prescribed reference posture and local energetic optimum. The horizontal origin is the reference used for kinematic reconstruction. At zero variance, the illustrative curves show the normalized quadratic model \(\Delta\overline P^*=(\mu_\ell^*)^2+c_1^*\mu_\ell^*\), where \(\mu_\ell^*=\mu_\ell/L\), \(c_1^*=C_1/(C_2L)\), and \(\Delta\overline P^*=(\mathcal{E}_{\mathrm{met}}/T-P_0)/(C_2L^2)\) for a length scale \(L>0\) and \(C_2>0\). When \(c_1^*=0\) (solid curve), the reference is stationary and minimizes the local model. When \(c_1^*=0.8\) (dashed curve), the same reference remains the measurement origin but the optimum is shifted to \(\mu_{\ell,\min}^*=-0.4\). Thus, symmetric offset trials can identify the direction of the local energetic slope and test whether the imposed posture coincides with the unconstrained optimum. Line styles and markers preserve the comparison in grayscale.}
\label{fig:reference-optimum}
\end{figure}

The distinction between a prescribed reference posture and an energetic optimum is experimentally consequential. The reference value \(\ell_0\) fixes the origin from which video-derived deviations are measured, but it need not coincide with a stationary point of the reduced metabolic power. A nonzero \(C_1\) means that the quasi-static equilibrium manifold crosses the prescribed reference with a nonzero energetic slope: small mean deviations in one direction increase power, whereas deviations in the opposite direction initially decrease it. Figure~\ref{fig:reference-optimum} illustrates this distinction. Symmetric positive and negative offsets can therefore determine not only whether the reference is stationary, but also on which side of it the local energetic minimum lies. The curvature coefficient \(C_2\) controls the displacement of that minimum and the energetic penalty for moving away from it.

\subsection{Dimensional consistency}

If \(\mathcal{E}_{\mathrm{met}}\) is measured in joules,
\(T\) in seconds, and \(\ell\) in metres, then
\[
[P_0]=\mathrm{W},
\qquad
[\Delta\mathcal{A}_{\ell}]=\mathrm{m\,s},
\qquad
[\mathcal{M}_{2,\ell}]=\mathrm{m^2\,s},
\]
and hence
\[
[C_1]=\mathrm{W\,m^{-1}},
\qquad
[C_2]=\mathrm{W\,m^{-2}}.
\]
These units provide a direct check on both analytical derivations and regression implementations.

\subsection{Video-based experimental identification}
\label{sec:video-identification}

The theoretical reduction can be implemented without resolving every microscopic contribution to muscle energetics. The required kinematic inputs are trial duration, mean length deviation, and deviation variance. A practical measurement sequence is summarized in Table~\ref{tab:video-protocol}.

\begin{table}[H]
\centering
\caption{Video-based identification sequence.}
\label{tab:video-protocol}
{\small
\renewcommand{\arraystretch}{1.15}
\begin{tabular}{p{0.15\textwidth}p{0.33\textwidth}p{0.40\textwidth}}
\hline
Stage & Operation & Output and purpose \\
\hline
Reference
&
Record the prescribed posture and quantify tracking uncertainty.
&
Reference length \(\ell_0\) and a calibration estimate of tracking variance.\\

Kinematics
&
Track the shoulder, elbow, and wrist, or another task-specific landmark set.
&
Joint trajectory \(\theta(t)\) reconstructed from the anatomical segment axes.\\

Geometry
&
Evaluate the selected mapping \(\ell(\theta)\) separately for every frame.
&
Length deviation \(x(t)=\ell(t)-\ell_0\). Framewise conversion preserves nonlinear geometric effects.\\

Moments
&
Average the reconstructed length deviations and their squared departures from the mean.
&
Video-derived predictors \(\widehat{\mu}_\ell\) and \(\widehat{\sigma}_\ell^2\).\\

Energetics
&
Combine the kinematic predictors with synchronized metabolic measurements.
&
Effective coefficients \(P_0\), \(C_1\), and \(C_2\).\\
\hline
\end{tabular}
}
\end{table}

The geometric conversion must precede averaging. In general,
\[
\overline{\ell(\theta)}
\neq
\ell(\overline{\theta}),
\]
because curvature of the angle--length relation allows angular variance to contribute to mean muscle length. The corresponding second-order correction and the discrete quadrature formulas are given in Appendix~\ref{app:video-moments}.

For trial \(j\), the reconstructed moments give
\begin{equation}\label{eq:video-moment-predictors}
\widehat{\Delta\mathcal{A}}_{\ell,j}
=
T_j\widehat{\mu}_{\ell,j},
\qquad
\widehat{\mathcal{M}}_{2,\ell,j}
=
T_j
\left(
\widehat{\mu}_{\ell,j}^{\,2}
+
\widehat{\sigma}_{\ell,j}^{\,2}
\right).
\end{equation}
The first expression shows the practical role of video-derived mean displacement: when \(C_1\neq0\), it supplies the absement correction \(C_1T_j\widehat{\mu}_{\ell,j}\). When \(C_1=0\), the leading kinematic correction is instead quadratic.

To keep the energy- and power-level formulations dimensionally equivalent, define the nuisance covariates \(\boldsymbol{z}_j\) and the participant effect \(u_{p(j)}\) on the power scale. For energy-level observations, the aggregate model is then
\begin{equation}\label{eq:energy-regression}
\mathcal{E}_j
=
P_0T_j
+
C_1\widehat{\Delta\mathcal{A}}_{\ell,j}
+
C_2\widehat{\mathcal{M}}_{2,\ell,j}
+
T_j
\left(
\boldsymbol{b}^{\mathsf T}\boldsymbol{z}_j
+
u_{p(j)}
\right)
+
e_j^{(E)}.
\end{equation}
Writing \(\overline P_j:=\mathcal{E}_j/T_j\) and dividing by \(T_j\) gives the power-normalized form
\begin{equation}\label{eq:power-regression}
\overline P_j
=
P_0
+
C_1\widehat{\mu}_{\ell,j}
+
C_2
\left(
\widehat{\mu}_{\ell,j}^{\,2}
+
\widehat{\sigma}_{\ell,j}^{\,2}
\right)
+
\boldsymbol{b}^{\mathsf T}\boldsymbol{z}_j
+
u_{p(j)}
+
e_j^{(P)},
\qquad
e_j^{(P)}
=
\frac{e_j^{(E)}}{T_j}.
\end{equation}
Here, \(\boldsymbol{z}_j\) contains prespecified covariates, while \(\boldsymbol{b}^{\mathsf T}\boldsymbol{z}_j\) and the participant-level random intercept \(u_{p(j)}\) are power contributions. If average power is measured directly rather than obtained by dividing an energy observation by duration, \(e_j^{(P)}\) is defined directly on the power scale rather than through the final identity in Eq.~\eqref{eq:power-regression}.

The design should vary mean displacement and variance as independently as practicable. Trials with approximately symmetric positive and negative offsets isolate the signed coefficient \(C_1\), whereas a zero-mean fluctuating condition provides an independent test of the \(C_2\sigma_\ell^2\) contribution. A compact symmetric design and its coefficient contrasts are given in Appendix~\ref{app:video-moments}.

The measured coordinate must retain a consistent physical meaning across trials. Muscle--tendon path length may be reconstructed from video-derived joint kinematics and subject-scaled musculoskeletal geometry, for example through an OpenSim workflow \cite{Delp2007}. Fascicle length may instead be obtained from synchronized ultrasound \cite{Kwah2013,Rosa2021}, but it should not be identified with muscle--tendon path length unless tendon compliance and pennation are incorporated explicitly.

Reference calibration is especially important because a constant length offset accumulates linearly in absement, whereas zero-mean tracking noise inflates the second moment. Joint moment, activation or electromyographic indices, co-contraction, and external load should also be recorded to test whether all trials sample the same reduced function \(\varphi\). Metabolic energy should be corrected for resting expenditure and evaluated over intervals compatible with the response characteristics of indirect calorimetry \cite{SelingerDonelan2014}.

Equations~\eqref{eq:energy-regression} and \eqref{eq:power-regression} are plug-in regressions: the hatted kinematic predictors are estimated rather than observed without error. Calibration-based correction of the second moment removes the leading additive noise contribution under the assumptions stated in Appendix~\ref{app:video-moments}, but it does not by itself eliminate attenuation or other errors-in-variables bias in the fitted coefficients. Predictor uncertainty should therefore be propagated by repeated calibration, resampling, or an explicit measurement-error model.

The regressions identify the effective coefficients \(P_0,C_1,C_2\). Without independent force, activation, and geometric measurements, they do not uniquely determine the individual derivatives of \(F_{\mathrm{tot}}\), \(\Pi\), \(Q\), or \(a_\ast\).

\section{Discussion}

The contribution of the present framework lies in connecting three elements that are usually treated separately: quasi-static moment balance, a local energetic reduction, and video-derived moment predictors. Once activation has been eliminated, Eq.~\eqref{eq:main-expansion} shows that the kinematic contribution is determined through second order by trial duration, mean length deviation, and the uncentred second moment. The expansion is therefore useful not merely as a formal Taylor series but as a compact and experimentally falsifiable measurement model.

Absement is the first raw integral moment of length deviation, rather than a generally sufficient physiological statistic. Its signed cancellation property is expected: opposite deviations cancel from the first moment while continuing to contribute to the second. Equation~\eqref{eq:raw-moment-decomposition} separates the quadratic contribution into squared mean displacement and variance, thereby preventing a sustained mean shift from being conflated with variability and preventing zero signed absement from being interpreted as absence of a kinematic energetic contribution.

The rearrangement invariance of the autonomous functional is useful in the intended quasi-static regime. Periodic, irregular, and phase-shifted profiles with the same empirical distribution of length values have the same model prediction. Consequently, a protocol need not estimate a separate energetic coefficient for every correction cycle: the mean offset supplies the absement predictor, and the dispersion supplies the quadratic predictor. Figure~\ref{fig:controlled-trajectories} illustrates this reduction for a sinusoidal deviation.

The framework is compatible with established muscle-energy models in which metabolic power depends on activation, force, and contractile state \cite{Russ2002,Umberger2003,Bhargava2004}. It does not imply universal frequency independence. Findings on rapid force production and varying isometric forces \cite{VanDerZeeKuo2021,Muralidhar2025} indicate that rate-dependent terms may be required outside the quasi-static range. The proposed video-based protocol therefore also supplies a direct model test: after controlling for \(\mu_\ell\) and \(\sigma_\ell^2\), a remaining systematic frequency dependence would mark the failure of the autonomous reduction.

\subsection{Limitations}

The model has several limitations:
\begin{itemize}
    \item \textbf{Quasi-static approximation.}
    Inertial, viscous, and rate-dependent muscle effects are omitted.

    \item \textbf{Local approximation.}
    The remainder is cubic only while the trajectory remains in a neighbourhood where the derivatives are bounded and the inverse coordinate map exists.

    \item \textbf{Scalar coordinate.}
    A single effective length cannot represent redistribution of load among synergists, antagonist co-contraction, or multi-joint coordination.

    \item \textbf{Interior activation.}
    The implicit reduction can fail near activation bounds or at points where \(Q_a=0\).

    \item \textbf{Instantaneous reduced power.}
    The functional assigns the same prediction to profiles with the same value distribution and duration. Rate, ordering, and spectral differences require additional variables.

    \item \textbf{Effective coefficients.}
    The regression coefficients combine geometry, force production, activation, and metabolic conversion. They should not be interpreted as isolated physiological constants.

    \item \textbf{Coordinate dependence.}
    Muscle--tendon path length and fascicle length describe different mechanical quantities. The fitted coefficients are meaningful only with respect to the coordinate used to obtain them.
\end{itemize}

\subsection{Extensions}

For a multi-joint or multi-muscle coordinate
\(\boldsymbol{x}(t)\in\mathbb{R}^n\), the corresponding local expansion is
\[
\mathcal{E}
=
P_0T
+
\boldsymbol{c}^{\mathsf T}
\int_0^T\boldsymbol{x}(t)\,dt
+
\frac{1}{2}
\int_0^T
\boldsymbol{x}(t)^{\mathsf T}
H\boldsymbol{x}(t)\,dt
+
O\!\left(T\|\boldsymbol{x}\|_{L^\infty}^3\right),
\]
where \(\boldsymbol{c}\) is the gradient and \(H\) is the Hessian of the reduced power at the reference posture. This formulation introduces a vector-valued deviation absement and includes cross-terms between coordinates.

A rate-dependent extension could instead begin from
\[
\mathcal{E}
=
\int_0^T
\Psi\bigl(\boldsymbol{x}(t),
\dot{\boldsymbol{x}}(t),
\boldsymbol{a}(t)\bigr)\,dt.
\]
Its expansion would contain velocity-dependent quadratic forms and could distinguish profiles that the present model treats as equivalent. Experimental comparison of the scalar model with such extensions would determine whether absement and displacement variance are adequate for the frequency range under study.

\section{Conclusions}

For a smooth quasi-static muscle model reduced to an effective length coordinate, total metabolic energy is controlled locally by trial duration and the first two raw integral moments of length deviation, as stated in Eq.~\eqref{eq:main-expansion}. Deviation absement equals duration multiplied by mean deviation and supplies the first-order kinematic correction when \(C_1\neq0\). If \(C_1=0\), that correction vanishes and the leading contribution is quadratic, with distinct components from squared mean displacement and variance.

A refined protocol can reconstruct \(\ell(t)\) from video-derived joint kinematics and use its first two moments directly in the energetic regression. Controlled positive and negative offsets identify the signed coefficient, while zero-mean fluctuating trials test the variance contribution and the quasi-static frequency range. A residual frequency effect after moment matching would provide a direct criterion for introducing rate-dependent variables.

\newpage

\appendix

\section{Consistency of the angle and length expansions}
\label{app:coordinate-conversion}

This appendix gives the second-order conversion between angular and length coordinates. It also identifies the correction that is lost if the inverse kinematic relation is truncated at first order.

Let
\[
y:=\theta-\theta_0,
\qquad
x:=\ell-\ell_0.
\]
From \eqref{eq:kinematic-expansion},
\begin{equation}\label{eq:appendix-forward}
x
=
s_0y
+
\frac{1}{2}\kappa_0y^2
+
O(|y|^3).
\end{equation}
Seek the inverse in the form
\[
y
=
b_1x+b_2x^2+O(|x|^3).
\]
Substitution into \eqref{eq:appendix-forward} gives
\[
x
=
s_0b_1x
+
\left(
s_0b_2+\frac{1}{2}\kappa_0b_1^2
\right)x^2
+
O(|x|^3).
\]
Matching coefficients yields
\[
b_1=\frac{1}{s_0},
\qquad
b_2=-\frac{\kappa_0}{2s_0^3}.
\]
Therefore,
\begin{equation}\label{eq:appendix-inverse}
\theta-\theta_0
=
\frac{\ell-\ell_0}{s_0}
-
\frac{\kappa_0}{2s_0^3}
(\ell-\ell_0)^2
+
O\!\left(|\ell-\ell_0|^3\right).
\end{equation}

Suppose that a reduced power has first been expanded in the angular coordinate:
\begin{equation}\label{eq:angular-power-expansion}
P(\theta_0+y)
=
P_0+p_1y+p_2y^2+O(|y|^3).
\end{equation}
Substitution of \eqref{eq:appendix-inverse} into
\eqref{eq:angular-power-expansion} gives
\[
P
=
P_0
+
\frac{p_1}{s_0}x
+
\left(
\frac{p_2}{s_0^2}
-
\frac{p_1\kappa_0}{2s_0^3}
\right)x^2
+
O(|x|^3).
\]
Hence the length-coordinate coefficients are
\begin{equation}\label{eq:coordinate-coefficients}
C_1
=
\frac{p_1}{s_0},
\qquad
C_2
=
\frac{p_2}{s_0^2}
-
\frac{p_1\kappa_0}{2s_0^3}.
\end{equation}
The second contribution to \(C_2\) is generated by the nonlinear inverse-coordinate transformation. Thus, the relation
\(C_2=p_2/s_0^2\) is valid only when
\(\kappa_0=0\) or \(p_1=0\).

The corresponding integral conversions are
\[
\int_0^T
\bigl(\theta(t)-\theta_0\bigr)\,dt
=
\frac{1}{s_0}\Delta\mathcal{A}_{\ell}
-
\frac{\kappa_0}{2s_0^3}
\mathcal{M}_{2,\ell}
+
O(T\varepsilon^3)
\]
and
\[
\int_0^T
\bigl(\theta(t)-\theta_0\bigr)^2\,dt
=
\frac{1}{s_0^2}\mathcal{M}_{2,\ell}
+
O(T\varepsilon^3).
\]
Substituting both relations into an angular energy expansion reproduces
\eqref{eq:coordinate-coefficients} and agrees with the direct reduced-function coefficients
\[
C_1=\varphi'(\ell_0),
\qquad
C_2=\frac{1}{2}\varphi''(\ell_0).
\]

\section{Explicit structure of the effective coefficients}
\label{app:explicit-coefficients}

This appendix shows how geometry, moment balance, and the metabolic-power law enter the aggregate coefficients. All derivatives of \(Q\) are evaluated at \((\theta_0,a_0)\), whereas derivatives of \(\Pi\) and \(F_{\mathrm{tot}}\) are evaluated at \((\ell_0,a_0)\).

Differentiating
\[
Q\bigl(\theta,a_\ast(\theta)\bigr)=0
\]
once and twice gives
\begin{align}
a_1
:=
a_\ast'(\theta_0)
&=
-\frac{Q_\theta}{Q_a},
\label{eq:a1}\\
a_2
:=
a_\ast''(\theta_0)
&=
-
\frac{
Q_{\theta\theta}
+
2Q_{\theta a}a_1
+
Q_{aa}a_1^2
}{
Q_a
}.
\label{eq:a2}
\end{align}
The first derivative describes the activation adjustment required by an infinitesimal angular displacement. The second derivative contains both the curvature of the equilibrium relation and the effect of the first-order activation response.

For the equilibrium function defined in the main text, the required partial derivatives are
\[
\begin{aligned}
Q_a
&=
\rho F_{\mathrm{tot},a},\\
Q_\theta
&=
\rho'F_{\mathrm{tot}}
+
\rho F_{\mathrm{tot},\ell}\ell_\theta
-
M_{\mathrm{ext}}',\\
Q_{\theta a}
&=
\rho'F_{\mathrm{tot},a}
+
\rho F_{\mathrm{tot},\ell a}\ell_\theta,\\
Q_{aa}
&=
\rho F_{\mathrm{tot},aa},\\
Q_{\theta\theta}
&=
\rho''F_{\mathrm{tot}}
+
2\rho'F_{\mathrm{tot},\ell}\ell_\theta\\
&\quad+
\rho
\left(
F_{\mathrm{tot},\ell\ell}\ell_\theta^2
+
F_{\mathrm{tot},\ell}\ell_{\theta\theta}
\right)
-
M_{\mathrm{ext}}''.
\end{aligned}
\]
These terms show explicitly that the equilibrium activation depends on the moment arm, the force--length relation, and the external moment, rather than on muscle length alone.

Conversion to the length coordinate gives
\begin{equation}\label{eq:ahat-derivatives}
\overline a_1
:=
\widehat a'(\ell_0)
=
\frac{a_1}{s_0},
\qquad
\overline a_2
:=
\widehat a''(\ell_0)
=
\frac{a_2}{s_0^2}
-
\frac{a_1\kappa_0}{s_0^3}.
\end{equation}
The term proportional to \(\kappa_0\) is the correction produced by the nonlinear angle--length transformation. Omitting it would generally give an incomplete quadratic coefficient.

Applying the chain rule to
\[
\varphi(\ell)
=
\Pi\bigl(\ell,\widehat a(\ell)\bigr)
\]
then yields
\begin{align}
C_1
&=
\Pi_\ell+\Pi_a\overline a_1,
\label{eq:C1-explicit}\\
C_2
&=
\frac{1}{2}
\left(
\Pi_{\ell\ell}
+
2\Pi_{\ell a}\overline a_1
+
\Pi_{aa}\overline a_1^2
+
\Pi_a\overline a_2
\right).
\label{eq:C2-explicit}
\end{align}
Thus, \(C_1\) measures the local power slope after the equilibrium activation has adjusted, whereas \(C_2\) contains the corresponding curvature and all second-order activation effects.

For the optional effective law~\eqref{eq:special-power-law}, the same result becomes
\[
\begin{aligned}
P_0
&=
\Pi_{\mathrm{b}}
+
\alpha a_0
+
\beta F_{\mathrm{act}}(\ell_0,a_0),\\
C_1
&=
\alpha\overline a_1
+
\beta
\left(
F_{\mathrm{act},\ell}
+
F_{\mathrm{act},a}\overline a_1
\right),\\
C_2
&=
\frac{\alpha}{2}\overline a_2\\
&\quad+
\frac{\beta}{2}
\left(
F_{\mathrm{act},\ell\ell}
+
2F_{\mathrm{act},\ell a}\overline a_1
+
F_{\mathrm{act},aa}\overline a_1^2
+
F_{\mathrm{act},a}\overline a_2
\right).
\end{aligned}
\]
This specialization is not required for regression of the aggregate coefficients, but it displays their physiological composition.

\section{Derivations supporting the video-based protocol}
\label{app:video-moments}

This appendix collects the technical results required by the experimental interpretation. The first part explains why moments, rather than individual cycles, are the natural predictors of the autonomous model. The remaining parts quantify finite-window effects, video discretization, geometric nonlinearity, measurement error, and coefficient identification.

\subsection{Occupation measure and moment hierarchy}

A recorded trial produces a large number of length values. For the autonomous functional, the relevant object is the fraction of the observation interval spent in each range of \(x=\ell-\ell_0\). This empirical distribution is represented by
\begin{equation}\label{eq:appendix-occupation-measure}
\nu_T(B)
=
\frac{1}{T}
\int_0^T
\mathbf{1}_B\bigl(x(t)\bigr)\,dt,
\qquad
m_n
=
\int_{\mathbb{R}}z^n\,d\nu_T(z)
=
\frac{\mathcal{M}_{n,\ell}}{T}.
\end{equation}
Here,
\[
\mathcal{M}_{n,\ell}
=
\int_0^T x(t)^n\,dt.
\]
The first moment is the mean deviation, and the second is the uncentred mean square.

The energy functional can be written as an average over this empirical distribution. More generally than required in the main theorem, if \(\varphi\in C^{p+1}\) in the relevant neighbourhood for some integer \(p\geq2\), Taylor expansion produces
\begin{equation}\label{eq:appendix-moment-hierarchy}
\begin{aligned}
\frac{\mathcal{E}_{\mathrm{met}}}{T}
&=
\int_{\mathbb{R}}
\varphi(\ell_0+z)\,d\nu_T(z),\\
\mathcal{E}_{\mathrm{met}}
&=
P_0T
+
\sum_{n=1}^{p}
C_n\mathcal{M}_{n,\ell}
+
R_{p+1},\\
C_n
&=
\frac{\varphi^{(n)}(\ell_0)}{n!},\\
|R_{p+1}|
&\leq
\frac{T}{(p+1)!}
\sup_{|z-\ell_0|\leq\varepsilon}
\left|
\varphi^{(p+1)}(z)
\right|
\varepsilon^{p+1}.
\end{aligned}
\end{equation}
Thus, absement and the quadratic integral are the first two members of a general moment hierarchy. The main text truncates this hierarchy after the second moment because the retained remainder is cubic.

\subsection{Periodic trajectories and finite observation windows}

Consider
\[
x(t)
=
\mu+A\sin(\omega t+\phi).
\]
The mean offset \(\mu\) and periodic amplitude \(A\) have different roles. Direct integration gives
\begin{align}
\Delta\mathcal{A}_{\ell}
&=
\mu T
+
\frac{A}{\omega}
\left[
\cos\phi-\cos(\omega T+\phi)
\right],
\label{eq:appendix-periodic-absement}\\
\mathcal{M}_{2,\ell}
&=
\mu^2T
+
\frac{2\mu A}{\omega}
\left[
\cos\phi-\cos(\omega T+\phi)
\right]
\nonumber\\
&\quad+
A^2
\left[
\frac{T}{2}
-
\frac{
\sin(2\omega T+2\phi)-\sin(2\phi)
}{4\omega}
\right].
\label{eq:appendix-periodic-second-moment}
\end{align}
The terms containing endpoint phases arise because a finite observation window need not contain a complete number of cycles. After division by \(T\), their relative importance decreases as \(1/(\omega T)\).

When
\[
\omega T=2\pi N,
\qquad
N\in\mathbb{N},
\]
the boundary terms vanish and
\begin{equation}\label{eq:appendix-complete-periods}
\Delta\mathcal{A}_{\ell}
=
\mu T,
\qquad
\mathcal{M}_{2,\ell}
=
T
\left(
\mu^2+\frac{A^2}{2}
\right).
\end{equation}
For the trajectory in Figure~\ref{fig:controlled-trajectories},
\(A=\sqrt{2}\sigma_\ell\), so the second expression becomes
\[
\mathcal{M}_{2,\ell}
=
T\bigl(\mu_\ell^2+\sigma_\ell^2\bigr).
\]

The disappearance of \(\omega\) from the complete-cycle result is a property of the autonomous quasi-static model, not a universal statement about muscle energetics. For example, a rate-dependent contribution proportional to
\(\int_0^T\dot{x}(t)^2\,dt\) would produce the mean value \(A^2\omega^2/2\) over complete cycles and would therefore distinguish frequencies directly.

\subsection{Discrete video estimators and geometric conversion}

The tracked landmarks determine \(\theta_k\) through the anatomical-axis construction in Eq.~\eqref{eq:joint-angle-geometry}. A calibrated geometric model is then applied to every frame:
\[
\ell_k=\ell(\theta_k),
\qquad
x_k=\ell_k-\ell_0.
\]
For frames
\[
t_0<t_1<\cdots<t_N,
\qquad
\Delta t_k=t_{k+1}-t_k,
\]
trapezoidal estimates of the two integral predictors are
\begin{equation}\label{eq:appendix-discrete-moments}
\begin{aligned}
\widehat{\Delta\mathcal{A}}_{\ell}
&=
\sum_{k=0}^{N-1}
\frac{\Delta t_k}{2}
\left(
x_k+x_{k+1}
\right),\\
\widehat{\mathcal{M}}_{2,\ell}
&=
\sum_{k=0}^{N-1}
\frac{\Delta t_k}{2}
\left(
x_k^2+x_{k+1}^2
\right),\\
\widehat{\mu}_\ell
&=
\frac{
\widehat{\Delta\mathcal{A}}_\ell
}{T_N},\\
\widehat{\sigma}_\ell^{\,2}
&=
\frac{
\widehat{\mathcal{M}}_{2,\ell}
}{T_N}
-
\widehat{\mu}_\ell^{\,2},
\qquad
T_N=t_N-t_0.
\end{aligned}
\end{equation}

Framewise geometric conversion is required because averaging and nonlinear coordinate transformation do not commute. Let
\[
y(t)=\theta(t)-\theta_0.
\]
Using the local expansion of \(\ell(\theta)\) and then averaging gives
\begin{equation}\label{eq:appendix-angle-mean-to-length}
\mu_\ell
=
s_0\mu_\theta
+
\frac{\kappa_0}{2}
\left(
\mu_\theta^2+\sigma_\theta^2
\right)
+
O\!\left(
\|y\|_{L^\infty}^3
\right).
\end{equation}
In particular, a zero mean angular deviation can still shift the mean length when \(\kappa_0\neq0\). Computing \(\ell(\theta_k)\) before averaging retains this contribution automatically.

\subsection{Reference bias and tracking noise}

For the discrete video record, let
\[
x_{\mathrm{obs},k}
=
x_k+b+\eta_k,
\]
where \(b\) is a constant reference bias. Conditional on the underlying trajectory, assume
\[
\mathbb{E}
\left[
\eta_k
\mid
x_0,\ldots,x_N
\right]
=0,
\qquad
\mathbb{E}
\left[
\eta_k^2
\mid
x_0,\ldots,x_N
\right]
=
\sigma_{\eta,k}^2.
\]
Let \(w_k\) denote the trapezoidal weight of frame \(k\), so that \(\sum_{k=0}^{N}w_k=T_N\). The conditional expectations of the observed estimators are then
\begin{equation}\label{eq:appendix-measurement-bias}
\begin{aligned}
\mathbb{E}
\left[
\widehat{\Delta\mathcal{A}}_{\mathrm{obs}}
\mid
x_0,\ldots,x_N
\right]
&=
\widehat{\Delta\mathcal{A}}_{\ell}
+
bT_N,\\
\mathbb{E}
\left[
\widehat{\mathcal{M}}_{2,\mathrm{obs}}
\mid
x_0,\ldots,x_N
\right]
&=
\widehat{\mathcal{M}}_{2,\ell}
+
2b\widehat{\Delta\mathcal{A}}_{\ell}
+
b^2T_N
+
\sum_{k=0}^{N}
w_k\sigma_{\eta,k}^2.
\end{aligned}
\end{equation}
For homoscedastic tracking noise, \(\sigma_{\eta,k}^2=\sigma_\eta^2\), the final term reduces to \(\sigma_\eta^2T_N\). The first relation shows why reference calibration is critical: even a small constant offset accumulates over the full trial. The second shows that conditionally zero-mean tracking noise does not cancel from the quadratic predictor.

If \(b\) is negligible after calibration and \(\sigma_{\eta,k}^2\) is estimated from a static calibration object or repeated reference images, the corrected second moment is
\begin{equation}\label{eq:appendix-corrected-second-moment}
\widehat{\mathcal{M}}_{2,\mathrm{corr}}
=
\widehat{\mathcal{M}}_{2,\mathrm{obs}}
-
\sum_{k=0}^{N}
w_k\widehat{\sigma}_{\eta,k}^2.
\end{equation}
Equations~\eqref{eq:appendix-measurement-bias} and \eqref{eq:appendix-corrected-second-moment} concern the discretized moments. Their difference from the continuous-time integrals is a separate quadrature error that should be checked by frame-rate sensitivity analysis. Moreover, correction of the mean second moment does not remove the errors-in-variables bias produced when noisy predictors are inserted into Eqs.~\eqref{eq:energy-regression} or \eqref{eq:power-regression}.

\subsection{Minimal symmetric identification design}

A compact calibration experiment can separate the signed and quadratic coefficients using four equal-duration conditions. Their second-order predictions are summarized in Table~\ref{tab:symmetric-design}.

\begin{table}[H]
\centering
\caption{Minimal symmetric identification design.}
\label{tab:symmetric-design}
{\small
\renewcommand{\arraystretch}{1.15}
\begin{tabular}{p{0.18\textwidth}p{0.15\textwidth}p{0.19\textwidth}p{0.30\textwidth}}
\hline
Condition
&
\(\mu_\ell\)
&
\(\sigma_\ell^2\)
&
\(\overline P-P_0\)\\
\hline
Reference
&
\(0\)
&
\(0\)
&
\(0\)\\

Positive offset
&
\(d\)
&
\(0\)
&
\(C_1d+C_2d^2\)\\

Negative offset
&
\(-d\)
&
\(0\)
&
\(-C_1d+C_2d^2\)\\

Zero-mean periodic
&
\(0\)
&
\(A^2/2\)
&
\(C_2A^2/2\)\\
\hline
\end{tabular}
}
\end{table}

The antisymmetric contrast between the positive and negative offsets isolates \(C_1\), whereas their symmetric part estimates \(C_2\). The periodic condition supplies an independent estimate of the same quadratic coefficient:
\begin{equation}\label{eq:appendix-identification-contrasts}
C_1
=
\frac{
\overline P_{+}-\overline P_{-}
}{2d},
\qquad
C_2
=
\frac{
\overline P_{+}
+
\overline P_{-}
-
2\overline P_0
}{2d^2}
=
\frac{
2\left(
\overline P_{\mathrm{per}}-\overline P_0
\right)
}{A^2}.
\end{equation}
In an actual experiment, nominal values \(d\) and \(A^2/2\) should be replaced by the video-derived \(\widehat{\mu}_\ell\) and \(\widehat{\sigma}_\ell^2\), and the complete regressions
\eqref{eq:energy-regression} or \eqref{eq:power-regression} should be used.

\bibliographystyle{plain}
\bibliography{references}

\end{document}